# Reservoir computing with dipole-coupled nanomagnets

Hikaru Nomura[1,2,3], Taishi Furuta[2], Kazuki Tsujimoto[3], Yuki Kuwabiraki[3],

Ferdinand Peper[4], Eiiti Tamura[2], Shinji Miwa[1,2,5], Minori Goto[1,2],

Ryoichi Nakatani[3], and Yoshishige Suzuki[1,2]

1. Center for Spintronics Research Network, Osaka University, Toyonaka, Osaka 560-8531, Japan

2. Graduate School of Engineering Science, Osaka University, Toyonaka, Osaka 560-8531, Japan

3. Graduate School of Engineering, Osaka University, Suita, Osaka 565-0871, Japan

4. Center for Information and Neural Networks, National Institute of Information and Communications Technology & Osaka University, Suita, Osaka 565-0871, Japan

5. The Institute for Solid State Physics, The University of Tokyo, Kashiwa Chiba 277-8581, Japan

Abstract

The feasibility of reservoir computing based on dipole-coupled nanomagnets is demonstrated using micro-magnetic simulations. The reservoir consists of an $2 \times 10$ array of nanomagnets. The static-magnetization directions of the nanomagnets are used as reservoir states. To update these states, we change the magnetization of one nanomagnet according to a single-bit-sequential signal. We also change the uniaxial anisotropy of the other nanomagnets using a voltage-induced magnetic-anisotropy change to enhance information flow, storage, and linear/nonlinear calculations. Binary tasks with AND, OR, and XOR operations were performed to evaluate the performance of the magnetic-array reservoir. The reservoir-computing output matrix was found to be trainable to perform AND, OR, and XOR operations with an input delay of up to three bits.

Recurrent neural networks (RNNs)[1] have been recognized as promising schemes for realizing high-performance artificial intelligence. Tremendous efforts have been made to realize RNNs using Si CMOS (complementary metal-oxide-semiconductor) technology and neuromorphic hardware based on different kinds of physical phenomena. Recently, neuromorphic hardware has attracted research attention because of its expected small size and small energy consumption. However, those trials to mimic the human brain seem to be still far from their goals. One obstacle facing realization of such hardware is the need for three-dimensional wiring that mimics the structure of the human brain.

In 2000, Cowburn and Welland introduced a magnetic quantum cellular automaton (MQCA)[2] composed of nanomagnets communicating wirelessly with one another. This interaction is achieved remotely via magnetic-dipole fields. In the MQCA, information is stored as the magnetization direction of the nanomagnets, whereas logic operations (changes of magnetization) are executed via the sum of the magnetic-dipole fields exerted from other nanomagnets. The MQCA can act as a transmission wire[2], but it is also possible to construct structures that operate like a majority logic gate[3, 4], shift register[5], or perform other tasks[6]. The MQCA has information non-volatility and binary-computing capability. Moreover, owing to the nanomagnets' nonlinear-magnetization process, the MQCA has the potential to calculate nonlinear functions.

Therefore, MQCAs are good candidates for realizing RNN hardware. It is, however, very difficult to implement learning algorithms on MQCAs, since the intensity of inter-magnetic interaction (synapse weight) is determined by the geometrical alignment of magnetic cells and is not tunable.

Meanwhile, reservoir computing[7-14] has appeared as a technique that may solve the problem of untunable interconnection weights. Reservoir computing comprises three parts: an input layer, a reservoir, and an output layer. The reservoir is an RNN with fixed interconnection weights. Even so, the reservoir itself has a high calculation power. The calculated results are distributed over the neurons in the reservoir, and are led through an output matrix thus producting output result in the output layer. The output matrix is adjusted through training using a teacher signal, and this allows the system to operate as an artificial neural network. Owing to this simple structure, most natural or artificial physical systems that contain nonlinear dynamics can be used as a reservoir. Recently, magnetic nano-oscillators have been utilized to recognize human voices within the framework of reservoir computing[15, 16]. This method, however, requires a high energy density, complicating its use as a high-density AI chip with low energy consumption.

In this study, we propose a new concept for reservoir computing using nanomagnets with magnetic-dipole interactions and external anisotropy control[7-14] (Fig. 1). As an example, a $2 \times 10$ array of nanomagnets [Fig. 1 (b)] is employed as a reservoir and its calculation power is demonstrated using micro-magnetic simulation.

The radii and thicknesses of the nanomagnets are 20 nm and 1 nm, respectively. The gaps between nanomagnets are 20 nm. The nanomagnet with an index of one in Fig. 1(b) is used for input. In this paper, input is limited to a single-bit-sequential binary signal, although one can treat analog inputs and outputs using the same magnetic array. At the beginning of the $k^{\text{th}}$ step, the reservoir receives the $k^{\text{th}}$ bit, $u_k$ in the sequential signal. For an input signal of the 0 (1), the input nanomagnet is magnetized toward the $+(-)z$ direction at beginning of the step.

Without bias voltage, the nanomagnets have perpendicular magnetic anisotropy. Therefore, in the absence of dipole interactions, each nanomagnet has a uniform magnetization perpendicular to the disk plane, i.e., either in the $+z$ or $-z$ directions. Because of its small size, the single domain state of each nanomagnet is assumed to be maintained, even during the dynamic process. Therefore, the nanomagnets are treated as macroscopic single magnetic dipole moments with constant vector length but different directions. Under magnetic-dipole interaction, the nanomagnets find their energy minima (which need not be the global minimum) by changing their magnetization direction dynamically.

The dynamics corresponding to the intelligent-calculation process are simulated by micro-magnetic simulation based on the Landau-Lifshitz-Gilbert (LLG) equation[17]. A single cell in the micro-magnetic simulation represents a nanomagnet with constant magnetic dipole moment in size and variable uniaxial anisotropy. The LLG equation at 0 K becomes

$$\frac{d\boldsymbol{M}_i}{dt} = -\gamma_{\text{LL}}\ \boldsymbol{M}_i \times \boldsymbol{H}_{\text{eff},i} - \frac{\alpha\gamma_{\text{LL}}}{M_s}\boldsymbol{M}_i \times (\boldsymbol{M}_i \times \boldsymbol{H}_{\text{eff},i}), \quad (1)$$

where

$$\boldsymbol{H}_{\text{eff},i} = \boldsymbol{H}_{\text{anisotropy},i} + \boldsymbol{H}_{\text{dipole},i}, \quad (2)$$

$$\boldsymbol{H}_{\text{anisotropy},i} = \frac{2}{\mu_0\ M_s^2}\begin{pmatrix} 0 & 0 & 0 \\ 0 & 0 & 0 \\ 0 & 0 & K_{\text{u},i}(t) \end{pmatrix} \cdot \boldsymbol{M}_i, \quad (3)$$

$$\boldsymbol{H}_{\text{dipole},i} = \sum_{j}^{N_{\text{mag}}} \boldsymbol{w}_{\text{dipole},ji} \cdot \boldsymbol{M}_j, \quad (4)$$

$$\boldsymbol{w}_{\text{dipole},ji} = \frac{1}{4\pi r_{ji}^5}\begin{pmatrix} 3r_{xji}^2 - r_{ji}^2 & 3r_{xji}r_{yji} & 3r_{xji}r_{zji} \\ 3r_{xji}r_{yji} & 3r_{yji}^2 - r_{ji}^2 & 3r_{yji}r_{zji} \\ 3r_{xji}r_{zji} & 3r_{yji}r_{zji} & 3r_{zji}^2 - r_{ji}^2 \end{pmatrix}. \quad (5)$$

Here, $N_{\text{mag}}$ is the number of nanomagnets, $i$ and $j$ are the indices of the nanomagnets $(i, j \in \{1, 2, \dots, N_{\text{mag}}\})$, $\boldsymbol{M}_i$ is the magnetic dipole moment of $i$-th magnet, $\boldsymbol{H}_{\text{eff},i}$ is the effective field, $\alpha$ is the damping constant, $\gamma_{\text{LL}} = \gamma/(1 + \alpha^2)$ is the gyromagnetic ratio in the Landau-Lifshitz form, $\gamma$ is the gyromagnetic ratio in the Landau-Lifshitz-Gilbert form, $\boldsymbol{H}_{\text{anisotropy},i}$ is the uniaxial-anisotropy field, $\boldsymbol{H}_{\text{dipole},i}$ is the dipole magnetic field, $K_{\text{u},i}(t)$ is the magnetic anisotropy energy $(K_{\text{u},i}(t) = 0$ or $K_{\text{u}0})$. $K_{\text{u}0}$ is the magnetic anisotropy energy under zero bias voltage. $\mu_0$ is the vacuum permeability, $M_s$ is the saturation magnetic dipole moment, $\boldsymbol{r}_{ji} = \{r_{xji}, r_{yji}, r_{zji}\}$ is the directional vector from the $j$-th to $i$-th nanomagnets. The parameters used here are as follows: $N_{\text{mag}} = 20$, $M_s = 1.3 \times 10^6$ A/m, $\gamma = 2.211 \times 10^5$ m/As, $K_{u0} = 0.1 \times \mu_0 M_s^2/V$ J, $V$ is volume of a nanomagnet and $\alpha = 0.5$. We solve the LLG equation using a fourth-order Runge-Kutta method.

As shown in Eqs. (2) to (5), all nanomagnets are connected by magnetic-dipole fields. The dynamics of a nanomagnet is determined by a weighted sum of the magnetizations, and the saturation property of magnetization results in a nonlinear operation on the magnetizations' weighted sum. Therefore, the system can be regarded as a finite-size realization of an infinite-range classical spin-glass (Hopfield) model[18, 19], although the weights are not randomly distributed in this case. In addition, the magnetizations are determined self-consistently. Therefore, the network is recurrent. The positive damping constant, $\alpha$, ensures the absence of auto-oscillation in the network. Therefore, the echo-state property[7] is preserved. Although the weights are fixed, the dynamics of the system is quite rich calculation inside in the neural network and is expected to act as an effective reservoir.

The operations of the system is divided insystem steps, each with seven stages, as shown in Fig. 2. In each step, only one bit of information is input to the system, so in order to receive a series of bits as input, the system trainsits through the corresponding steps. In each stage, we apply a voltage to selected nanomagnets to facilitate the flow of information, the storage of intermediate results, and control of nonlinear calculations. To select the magnets, we classify nanomagnets into Groups I to III according to the way in which the $K_\mathrm{u}$ values change [See Fig. 1(b)]. In the first stage $(p = 1)$ in each step, the voltage is not applied (Fig. 2), so all nanomagnets have perpendicular anisotropy. In the second stage $(p = 2)$ in each step, the $K_\mathrm{anisotropy}$ values of the nanomagnets in Groups II and III are changed to zero. Under this configuration, the magnetizations of the nanomagnets in Groups II and III are easily rotated by stray fields from the other nanomagnets, and the information stored in Group I is transferred to Groups II and III by partially incorporating information from the other nanomagnets. We set the time interval of the stages to be longer than the relaxation time of the magnetization dynamics. Therefore, at the end of each stage, the magnetizations align parallel to the self-consistent effective field:

$$\boldsymbol{M}_i \parallel \boldsymbol{H}_{\mathrm{eff},i} \ . \tag{6}$$

In the third stage $(p = 3)$ of each step, we fix the magnetizations of the nanomagnets in Group II so as to be close to either the $+z$ or $-z$ directions by changing their $K_u$s to $K_{u0}$. This is a nonlinear operation on the sum of the field. In the fourth stage ($p = 4$) of each step, the $K_u$ values of the nanomagnets in Group I are changed to zero. This stage is similar to the second stage but with a shift of 1 bit toward the $+y$ direction [see Fig. 1 (b)]. By repeating this process, we reach the first stage of the next step. Here the information stored in the nanomagnets is partially transferred along the $y$-axis direction by 3-dots, incorporating information from the other nanomagnets in a nonlinear way. In other words, the input signal information is slightly lost. This property should determine the short-term memory of the system[7]. The nanomagnets with even indices help communication between nanomagnets over a distance, enhancing the performance of the correlation calculation among multiple bits.

As a reservoir state $\boldsymbol{x}_k$, we use the $x$-component of the normalized magnetization $\mu_{xi}$ at the end of the third stage ($p = 3$) in each step $k$. $\mu_{xi}$ can be read by using magneto-resistive effects. The reservoir state at step $k$ then becomes

$$\mu_{xi}(k) = \frac{1}{2}\frac{M_{xi}(k)}{M_s} + \frac{1}{2}, \tag{7}$$

$$\boldsymbol{x}_k = \left\{\mu_{x1}(k), \mu_{x2}(k), \ldots, \mu_{xN_{\mathrm{mag}}}(k), 1\right\}. \tag{8}$$

Here, "1" as the final element on the right-hand side of Eq. (8) serves to adjust the mean output value of the reservoir computation, $o_k$, which is explained later. In an actual device, noise exists in the magnetization measurements, such that the number of significant figures for these measurements is limited. Therefore, we round the $x$-component of the magnetizations to no more than three significant figures when we testing the device.

The output, $o_k$, of the magnetic-array-reservoir computation then becomes

$$o_k = \boldsymbol{x}_k \cdot \boldsymbol{w}_f, \tag{9}$$

where $\boldsymbol{w}_f$ is the output vector of the reservoir computation and $f$ specifies a target function, i.e., the AND operation between successive input bits, for example. $\boldsymbol{w}_f$ is trained to minimize the square error between the teacher's answer and the output of the reservoir computation $\left(\sum_k^{N_{\text{train}}} (f - o_k)^2\right)$. Here, $f$ is a teacher's answer for a given input series.

To test the performance of the magnetic reservoir, we employed uniformly distributed random binary bits for the input, $u_k$. As target functions, we have chosen the AND, OR, and XOR operations between two bits with an $n$-step delay. The target functions are denoted as $\text{AND}(u_k, u_{k-n})$, $\text{OR}(u_k, u_{k-n})$, and $\text{XOR}(u_k, u_{k-n})$. Note that at step $k$, we only provide one bit of information, $u_k$. Therefore, the target functions require a short-term memory of $n$ steps.

Fig. 3 shows the output values of these functions. The white and red circles denote the values 0 and 1, respectively. Here, we consider how to classify the output values based on the input values $A$ and $B$. The outputs of the AND/OR functions are linearly separable (e.g., $A + B = 1.25$ and $A + B = 0.25$). On the other hand, the XOR outputs can only be separated by a nonlinear function (e.g., $4A^2 - 5A + 2AB - 4B + 3B^2 = -1$). Nonlinear-classification tasks like the XOR-function task are considered to be more difficult than linear ones.

In this study, 100 binary-input data samples ($N_{\text{train}} = 100$) were used for training. Fig. 4 shows typical values for the input data, the output of the teacher's answer, and the output of the magnetic-array reservoir for functions with various input delays, $n$, after the training. Comparison of the data in Fig. 4 shows that the reservoir-computing output with trained output vectors shows good agreement with the teacher's answers up to an input delay of three steps.

Here, we define the error rate between the teacher's answer $f$ and the trained reservoir-computing output $o_k$ as $\frac{1}{N_{\text{test}}}\sum_{k}^{N_{\text{test}}}|f - o_k|$, where $N_{\text{test}}$ is the number of binary-input data points used to test the error. We measure the error rates of four sets of 1,000 binary-input data points ($N_{\text{test}} = 1,000$). Up to an input delay of three steps ($n = 3$), the error rate is almost zero (less than the precision of the numerical simulation). On the other hand, the error rates of the AND/OR and XOR functions with input delays of four become $0.253 \pm 0.009$, $0.249 \pm 0.009$, and $0.502 \pm 0.007$, respectively. If reservoir computing yields 0 or 1 randomly, the error rate should be 0.5 for all functions.

On the other hand, if the reservoir recognizes only the first bit, the error rates for the AND, OR, and XOR functions will be 0.25, 0.25, and 0.5, respectively. The observed result matches the latter case. These results suggest that the nanomagnetic-array reservoir shown in Fig. 1(b) can perform AND, OR, and XOR functions up to an input delay of three steps.

In our design, three rows (one data row and two buffer rows) are required to perform a one-bit shift operation. Therefore, an array with $10 (= 3 \times 3 + 1)$ rows can perform four-bit shift operations and data older than three steps will be lost. Thus, the short-term memory is limited to three steps (bits). By increasing the number of elements in the row, we can increase the upper limit of the short-term memory. However, the maximum short-term memory will depend on the interference strength with even-number dots. The operation between two bits with a longer delay requires higher accuracy for the magnetization measurements in proportion to the increment of the distances. This means that the performance of reservoir computing is limited not only by the number of rows, but also by the number of significant figures.

Here, we only demonstrated a magnetic array that composes a square lattice. However, in reservoir computing, the connection strengths between the nodes are generally chosen to be random in order to perform general tasks. A randomly arranged nanomagnet array may have the potential to act as a better reservoir. Since it is not straightforward to predict the performance of larger-scale arrays, further theoretical investigation is required. Beyond this, the production of larger-scale hardware is feasible using recent technology to fabricate magnetoresistive-random-access memories (MRAMs).

In this letter, we introduced and demonstrated reservoir computing based on nanomagnets with magnetic interactions. Sequential single-bit inputs cause rearrangement of the magnetizations of the nanomagnets. The dynamics include intelligent calculation. In addition, it has been shown that intentional control of magnetic anisotropy using voltage applications may control information flow, storage, and nonlinear calculations. Using such techniques, the magnetic-array reservoir could be trained to perform AND, OR, and XOR functions with input delays of up to three steps. The proposed hardware can be realized and even enlarged in array size by using recent technology to fabricate MRAMs.

This research and development work was supported by the Ministry of Internal Affairs and Communications, JAPAN.

References

1. H. T. Siegelmann and E. D. Sontag, Appl. Math. Lett. **4** (6), 77-80 (1991).
2. R. P. Cowburn and M. E. Welland, Science **287** (5457), 1466-1468 (2000).
3. A. Imre, G. Csaba, L. Ji, A. Orlov, G. H. Bernstein and W. Porod, Science **311** (5758), 205-208 (2006).
4. N. Hikaru and N. Ryoichi, Appl. Phys. Express **4** (1), 013004 (2011).
5. N. Hikaru, Y. Naomichi, M. Soichiro and N. Ryoichi, Appl. Phys. Express **10** (12), 123004 (2017).
6. A. Orlov, A. Imre, G. Csaba, L. Ji, W. Porod and G. H. Bernstein, J. Nanoelectron. Optoe. **3** (1), 55-68 (2008).
7. H. Jaeger, GMD Report, 148 (2001).
8. W. Maass, T. Natschlager and H. Markram, Neural. Comput. **14** (11), 2531-2560 (2002).
9. H. Jaeger and H. Haas, Science **304** (5667), 78-80 (2004).
10. D. Verstraeten, B. Schrauwen, M. D'Haene and D. Stroobandt, Neural. Netw. **20** (3), 391-403 (2007).
11. L. Appeltant, M. C. Soriano, G. Van der Sande, J. Danckaert, S. Massar, J. Dambre, B. Schrauwen, C. R. Mirasso and I. Fischer, Nat. Commun. **2**, 468 (2011).
12. K. Vandoorne, P. Mechet, T. Van Vaerenbergh, M. Fiers, G. Morthier, D. Verstraeten, B. Schrauwen, J. Dambre and P. Bienstman, Nat. Commun. **5**, 3541 (2014).
13. K. Nakajima, H. Hauser, T. Li and R. Pfeifer, Sci. Rep-Uk. **5**, 10487 (2015).
14. K. Fujii and K. Nakajima, Phys. Rev. Appl. **8** (2), 024030 (2017).
15. J. Torrejon, M. Riou, F. A. Araujo, S. Tsunegi, G. Khalsa, D. Querlioz, P. Bortolotti, V. Cros, K. Yakushiji, A. Fukushima, H. Kubota, S. Y. Uasa, M. D. Stiles and J. Grollier, Nature **547** (7664), 428 (2017).
16. T. Furuta, K. Fujii, K. Nakajima, S. Tsunegi, H. Kubota, Y. Suzuki and S. Miwa, Phys. Rev. Appl. **10** (3), 034063 (2018).
17. T. L. Gilbert, IEEE Trans. Magn. **40** (6), 3443-3449 (2004).
18. J. J. Hopfield, Proc. Natl. Acad. Sci-Biol. **79** (8), 2554-2558 (1982).
19. D. J. Amit and H. Gutfreund, Phys. Rev. A **32** (2), 1007-1018 (1985).

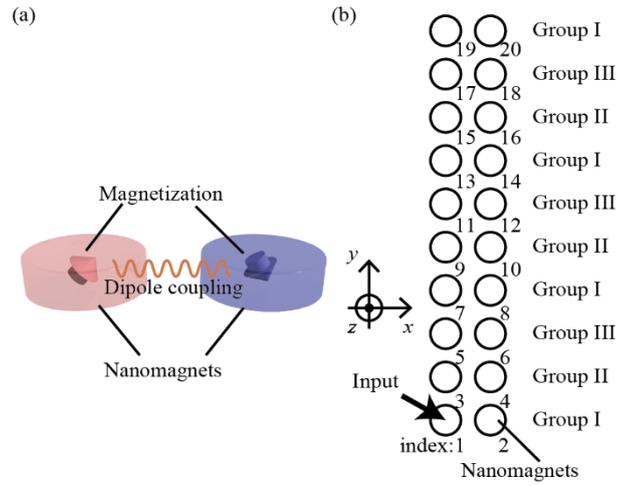

Fig. 1. (a) Schematic illustration of dipole-coupled nanomagnets and (b) schematic top view of a nanomagnetic-array-based reservoir. The numbers shown at the lower right of the dots are their indices.

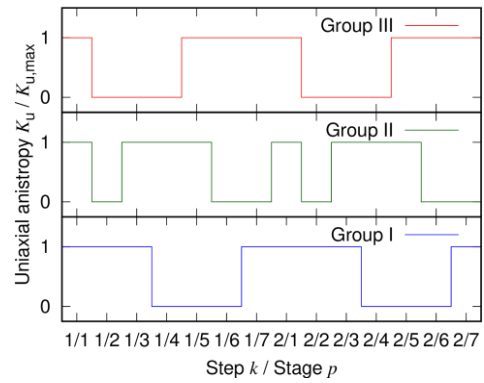

Fig. 2. Update scheme of the reservoir state, in which only the first two steps are shown. The uniaxial-anisotropy constants of the nanomagnets are modified at each stage.

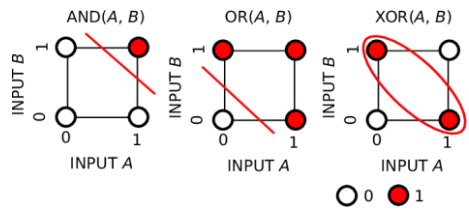

Fig. 3. Outputs of the AND, OR, and XOR functions. The outputs of each function can be divided into different classes with red lines. The white and red circles denote output values of 0 and 1, respectively.

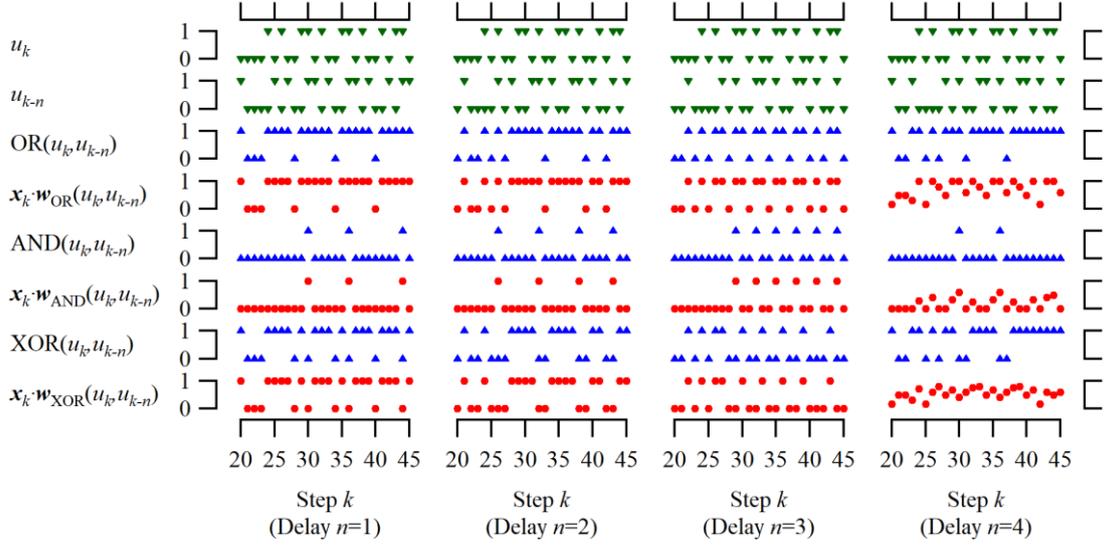

Fig. 4. Typical values of input data $u_k$, teacher functions $f(u_k, u_{k-n})$, and the nanomagnetic-array-reservoir output $o_k = \boldsymbol{x} \cdot \boldsymbol{w}_{f(u_k, u_{k-n})}$, where $n$ is the input delay for the teacher function.